\documentclass[reprint,twocolumn,superscriptaddress,amsmath,amssymb,jcp]{revtex4-1}
\usepackage{dcolumn}
\usepackage{epsfig}
\usepackage{graphics}
\usepackage[latin1]{inputenc} 
\usepackage[T1]{fontenc}
\usepackage{bm}
\usepackage{amssymb}
\newcommand{\ddst}{false}

\begin{document}

\title{Fracture toughness of calcium--silicate--hydrate grains from molecular dynamics simulations}

\author{M. Bauchy}
 \email[Contact: ]{bauchy@ucla.edu}
 \homepage[\\Homepage: ]{http://mathieu.bauchy.com}
 \affiliation{Department of Civil and Environmental Engineering, University of California, Los Angeles, CA 90095, United States}
\author{H. Laubie}
 \affiliation{Concrete Sustainability Hub, Department of Civil and Environmental Engineering, Massachusetts Institute of Technology, 77 Massachusetts Avenue, Cambridge, MA 02139, United State}
\author{M. J. Abdolhosseini Qomi}
 \affiliation{Concrete Sustainability Hub, Department of Civil and Environmental Engineering, Massachusetts Institute of Technology, 77 Massachusetts Avenue, Cambridge, MA 02139, United State}
\author{C. G. Hoover}
 \affiliation{Concrete Sustainability Hub, Department of Civil and Environmental Engineering, Massachusetts Institute of Technology, 77 Massachusetts Avenue, Cambridge, MA 02139, United State}
\author{F.-J. Ulm}
 \affiliation{Concrete Sustainability Hub, Department of Civil and Environmental Engineering, Massachusetts Institute of Technology, 77 Massachusetts Avenue, Cambridge, MA 02139, United State}
\author{R. J.-M. Pellenq}
 \affiliation{Concrete Sustainability Hub, Department of Civil and Environmental Engineering, Massachusetts Institute of Technology, 77 Massachusetts Avenue, Cambridge, MA 02139, United State}
 \affiliation{MIT-CNRS joint laboratory at Massachusetts Institute of Technology, 77 Massachusetts Avenue, Cambridge, MA 02139, United States}
 \affiliation{Centre Interdisciplinaire des Nanosciences de Marseille, CNRS and Aix-Marseille University, Campus de Luminy, Marseille, 13288 Cedex 09, France}

\date{\today}

\begin{abstract}
Cement is the most widely used manufacturing material in the world and improving its toughness would allow for the design of slender infrastructure, requiring less material. To this end, we investigate by means of molecular dynamics simulations the fracture of calcium--silicate--hydrate (C--S--H), the binding phase of cement, responsible for its mechanical properties. For the first time, we report values of the fracture toughness, critical energy release rate, and surface energy of C--S--H grains. This allows us to discuss the brittleness of the material at the atomic scale. We show that, at this scale, C--S--H breaks in a ductile way, which prevents from using methods based on linear elastic fracture mechanics. Knowledge of the fracture properties of C--S--H at the nanoscale opens the way for an upscaling approach to the design of tougher cement.
\end{abstract}

\maketitle

\section{Introduction}
\label{sec:intro}

Cement is the most used material in the world \cite{scrivener_straight_2012}. Thanks to its low cost, it is the only material that can satisfy the growing demand for infrastructure, especially in developing countries. However, the production of cement is responsible for about 7\% of the global emissions of carbon dioxide in the atmosphere \cite{mehta_reducing_2001}. Because of such a ubiquitous presence in our environment, only a small decrease in its production would have a significant impact in terms of greenhouse gas emissions. To this end, one option is to improve the toughness of cement. Indeed, tougher cement would allow using less material while achieving comparable mechanical properties. Moreover, an increased resistance to fracture would improve its longevity, making it more sustainable.

Due to its multi-scale \cite{masoero_nanostructure_2012} and heterogeneous \cite{allen_composition_2007} nature, understanding the fracture mechanism of calcium--silicate--hydrate (C--S--H), the binding phase of cement, remains challenging. In particular, the intrinsic fracture toughness of C--S--H grains at the nanoscale remains unknown, and it would be challenging to obtain it experimentally. This knowledge would serve as a basis to build a multi-scale model of fracture in C--S--H, following a bottom-up approach. Despite the prevalence of cement in the built environment, the molecular structure of C--S--H has just recently been proposed \cite{pellenq_realistic_2009, manzano_confined_2012, abdolhosseini_qomi_concrete_2013, abdolhosseini_qomi_applying_2013, bauchy_order_2014}, which makes it possible to investigate its mechanical properties at the nanoscale.

Hence, relying on this newly available model, we computed the fracture toughness and critical energy release rate of C--S--H at the nanoscale by means of molecular dynamics simulations. On the other hand, the computation of its surface energy allowed us to quantify its brittleness. This paper is organized as follows. We first present the details of the simulation of C--S--H in Sec. \ref{sec:simu}, as well as the methodology used to obtain the fracture toughness in Sec. \ref{sec:met}. Results are reported in Sec. \ref{sec:res} and discussed in Sec. \ref{sec:dis}. Some conclusions are finally presented in Sec. \ref{sec:ccl}.

\section{Simulation details}
\label{sec:simu}

To describe the disordered molecular structure of C--S--H, Pellenq et al. \cite{pellenq_realistic_2009} proposed a realistic model for C--S--H with the stoichiometry of (CaO)$_{1.65}$(SiO$_2$)(H$_2$O)$_{1.73}$. We generated the C--S--H model by introducing defects in an 11 \AA\ tobermorite \cite{hamid_crystal-structure_1981} configuration, following a combinatorial procedure. 11 \AA\ tobermorite consists of pseudo-octahedral calcium oxide sheets, which are surrounded by silicate tetrahedral chains. The latter consists of bridging oxygen atoms and $Q^2$ silicon atoms (having two bridging and two non-bridging terminal oxygen atoms) \cite{abdolhosseini_qomi_evidence_2012}. Those negatively charged calcium--silicate sheets are separated from each other by an interlayer spacing, which contains water molecules and charge-balancing calcium cations. Whereas the Ca/Si ratio in 11 \AA\ tobermorite is 1, this ratio is increased to 1.71 in the present C--S--H model, through randomly removing SiO$_2$ groups. Defects in silicate chains provide possible sites for adsorption of extra water molecules. The adsorption of water molecules in the structurally defected tobermorite model was performed via the Grand Canonical Monte Carlo method, ensuring equilibrium with bulk water at constant volume and room temperature. The REAXFF potential \cite{manzano_confined_2012}, a reactive potential, was then used to account for the reaction of the interlayer water with the defective calcium--silicate sheets. The use of the reactive potential allows observing the dissociation of water molecules into hydroxyl groups. More details on the preparation of the model and its experimental validation can be found elsewhere \cite{pellenq_realistic_2009,abdolhosseini_qomi_concrete_2013,qomi_anomalous_2014}.

\begin{figure}
\includegraphics*[width=\linewidth, keepaspectratio=true, draft=\ddst]{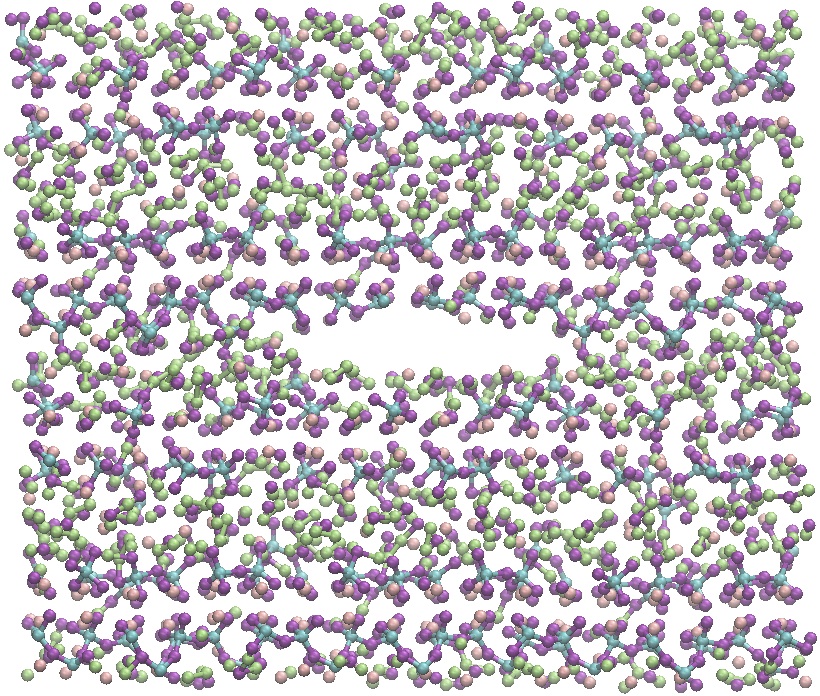}
\caption{\label{fig:snap} (Color online) Snapshot of the atomic configuration of the calcium--silicate--hydrate model, with an initial crack. Silicon, oxygen, calcium, and hydrogen atoms are respectively represented in cyan, purple, pink, and green.}
\end{figure}

We simulated the previously presented C--S--H model, made of 501 atoms, by molecular dynamics using the LAMMPS package \cite{plimpton_fast_1995}. To this end, we used the REAXFF potential \cite{manzano_confined_2012} with a time step of 0.25fs. We first relaxed the system at zero pressure and 300K during 2.5 ns in the NPT and NVT ensembles and made sure that convergence of the energy and volume was achieved.  To study the propagation of an initial crack in the material, the initial cell was replicated in $x$, $y$ and $z$ direction. The smallest considered system was made of 1x2x2 initial cells, whereas the biggest one was built by a 2x5x3 replication. Fig. \ref{fig:snap} shows a snapshot of the atomic configuration of a 1x3x2 C--S--H model. After the replication, each system was relaxed during 1 ns in NPT and NVT ensembles.

\section{Methodology}
\label{sec:met}

Traditional methods of determining fracture energy, for example using the classic path independent J-integral evaluation either analytically \cite{rice_mathematical_1968, rice_path_1968} or numerically \cite{smilauer_multiscale_2011}, will not work for this investigation because they assume a large specimen made of a continuous homogeneous material, such that the stresses emanating from and surrounding the crack tip are uninfluenced by structural boundaries. Other models such as the so-called \textit{Work of Fracture} method or \textit{fictitious crack model} of Hillerborg \cite{hillerborg_theoretical_1985} and enshrined in a RILEM international recommendation \cite{_determination_1985}, which relates the external work to the internal energy release rate, (similar proposals were made for ceramics \cite{nakayama_direct_1965, tattersall_work_1966}), are also not applicable for the same reasons. Other methods such as utilizing the linear elastic fracture mechanics (LEFM) energy release rate functions for known geometries \cite{tada_analysis_2000} or approximations for more complicated geometries using the Jenq-Shah model \cite{jenq_fracture_1985, jenq_two_1985}, which can also yield crack extension, are not applicable for this type of analysis. Therefore, another method must be adopted.

Recently, Brochard et al. \cite{brochard_fracture_2013} introduced a new method to study fracture properties at the smallest scales, based on molecular dynamics simulations. This approach relies on the energetic theory of fracture mechanics \cite{griffith_phenomena_1921, leblond_mecanique_2003, anderson_fracture_2005} and consists of thermodynamic integration during crack propagation. This method does not involve any assumption about the mechanical behavior of the material during the fracture and can thus capture fracture properties of brittle as well as ductile systems \cite{brochard_fracture_2013}. 

In the following, we focus on fracture in mode I, i.e., with an opening mode and a loading normal to the crack plane. A crack is first initiated into a molecular sample. We restrict ourselves to the propagation of cracks in planes parallel to the calcium layers, that is, the direction of fracture along the $z$ axis being the weakest, as it does not involve any bond breaking inside the silicate chains. We also focus on crack propagation inside interlayer spaces, where both structural and free water can be found, as it is the weakest point of rupture in the system. Such cracks are expected to exist naturally in real materials and allow us to control in which interlayer space the crack will propagate. The initial crack is created by removing atoms located inside an elliptic volume along the $x$ direction. The ellipse is chosen to be five times larger in the $y$ direction than in the $z$ direction, thus inducing a strong concentration of the stress at the crack tips. Its length goes from 12 to 50 \AA, depending on the size of the supercell. Note that the initial length must be long enough for the initial hole to be stable but small as compared with the box length in the $y$ direction.

Before any tension is applied, the system is fully relaxed to be unstressed; thus, its mechanical energy $P$, involved by strain, becomes zero. The procedure then consists in increasing the size $L_z$ of the system in the direction orthogonal to the initial crack until its full propagation along the $y$ axis. $L_z$ is  incremented stepwise by 1\% of its initial unstressed value $L_{z0}$ up to $L_{z {\rm max}}=1.5L_{z0}$. After each increase of the tensile strain $\epsilon = (L_z-L_{z0})/L_{z0}$, the system is relaxed for 5 ps before performing a statistical averaging stage for another 5 ps. During the latter phase, the stress in the $z$ direction $\sigma_z$ is computed with the virial equation \cite{allen_computer_1987}.

Note that the entire fracture simulation is operated within the canonical NVT ensemble, in which the temperature is controlled by a Nose--Hoover thermostat \cite{nose_molecular_1984,hoover_canonical_1985}. Hence, we are unable to capture potential heat transfers during the fracture. In fact, this procedure has not been designed to model the kinetics of crack propagation. On the contrary, thermodynamic quantities are always integrated when the system is at equilibrium, at each strain step. The phonons that arise during the fracture are annealed by the thermostat, as it will be shown later (see Sec. \ref{sec:res}). Therefore, phonons are not included in the following thermodynamic integration.

As the crack starts to propagate, some elastic energy $P$ is released to create new surface. This is captured by the energy release rate $G$:

\begin{equation}
 G = - \frac{\partial P}{\partial A}
\end{equation} where $A$ is the crack area. When propagation occurs, the energy release rate is equal to the critical energy release rate $G_c$, which is considered as a property of the material. Once the crack propagation is complete, the system becomes unstressed again, so that $P=0$, the mechanical energy having been released by crack propagation. The integration of $\sigma_z$ over the whole process, i.e., the external work, thus provides the critical energy release rate $G_c$:

\begin{equation}
 \label{eq:integration}
 G_c = \frac{\Delta F}{\Delta A} = \frac{L_x L_y}{\Delta A_{\infty}} \int _{L_{z0}}^{L_{z {\rm max}}} \sigma_z dL_z
\end{equation} where $F$ is the free energy of the system and $\Delta A_{\infty} = A_{\infty} - A_0$ is the total area of surface created at the end of the fracture, when the crack has fully propagated. This formula is a direct consequence of Griffith theory of fracture \cite{griffith_phenomena_1921}. It is worth noting that evaluating the crack area at the end of the fracture may not be straightforward as the created surface may show some roughness. To make an accurate estimate of the critical energy release rate, the real surface area has been calculated using the procedure proposed in Ref. \cite{brochard_fracture_2013}.

Alternatively to the energetic approach, the notion of fracture toughness $K_{Ic}$ is usually used in engineering application. This quantity was introduced by Irwin \cite{irwin_fracture_1958} as the maximum stress intensity at the crack tip a solid can undergo, and below which propagation cannot occur. The relationship between $K_{Ic}$ and $G_c$ is given by the Irwin formula \cite{irwin_fracture_1958}:

\begin{equation}
 G_c =  {\cal H}_I K_{Ic}^2
 \label{eq:irwin}
\end{equation} where ${\cal H}_I$ is given in Ref. \cite{sih_cracks_1965} for transversely isotropic solids and can be written in terms of the stiffness constants $C_{ij}$, using Voigt notation, as:

\begin{equation}
 {\cal H}_I = \frac{1}{2} \sqrt{\frac{C_{11}}{C_{11} C_{33}-C_{13}^2} \left( \frac{1}{C_{44}} + \frac{2}{C_{13}+\sqrt{C_{11} C_{33}}} \right)}
\end{equation} in plane strain, as is the case of the current study. Note that, although we rely on a general energetic approach that does not assume a purely brittle fracture, we keep in mind that the relation between $G_c$ and $K_{Ic}$ was derived in the context of LEFM. The full elastic tensor $C_{ij}$ was computed for a bulk system, before the introduction of the initial crack. The elements of the stiffness tensor are obtained by calculating the curvature of the potential energy $U$ with respect to small strain deformations $\epsilon_i$ \cite{pedone_insight_2007}:

\begin{equation}
C_{ij} =  \frac{1}{V} \frac{\partial^2 U}{\partial \epsilon_i \partial \epsilon_j}
\end{equation} where $V$ is the volume of the system. In isotropic materials, Eq.\ref{eq:irwin}  reduces to the usual Irwin formula \cite{leblond_mecanique_2003}:

\begin{equation}
 G_c =  \frac{1-\nu ^2}{E} K_{Ic}^2
\end{equation} where $E$ is the Young's modulus. However, C--S--H is not an isotropic material, as the Young's modulus in the direction perpendicular to the calcium layers is lower than that in the plane of the layers \cite{abdolhosseini_qomi_concrete_2013}.

\section{Results}
\label{sec:res}

\begin{figure}
\includegraphics*[width=\linewidth, keepaspectratio=true, draft=\ddst]{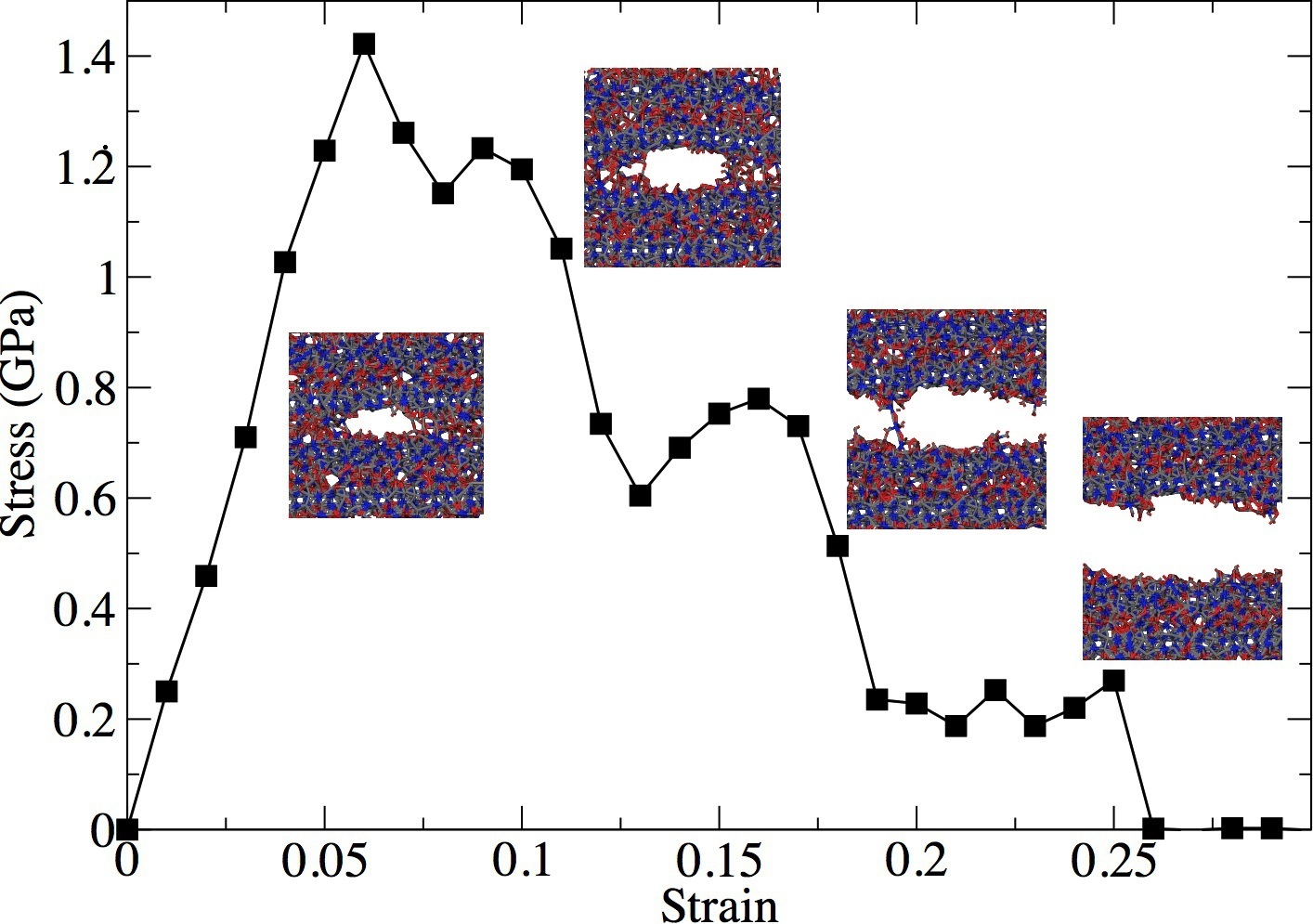}
\caption{\label{fig:fracture} (Color online) Computed stress as a function of the tensile strain imposed to the system. The snapshots show the molecular configurations at different stages of the fracture.}
\end{figure}

Fig. \ref{fig:fracture} shows the computed stress $\sigma_z$ with respect to the tensile strain $\epsilon$ for a C--S--H sample. The simulated system is 13.1 \AA\ $\times$ 54.4 \AA\ $\times$46.8 \AA\ in volume, with an initial crack of 15 \AA. At low strain (up to 6\%), the mechanical response is linear elastic. The stress thus increases linearly with the strain up to 1.4 GPa, the slope being related to the Young's modulus of the system. During this stage, the crack does not propagate and the free energy of the system is stored in the form of mechanical elastic energy only. At larger strain, the crack starts to propagate. Contrary to brittle materials like quartz \cite{brochard_fracture_2013}, C--S--H shows a strong ductile behavior in the sense that the crack does not propagate instantly after a given critical strain. Thanks to its internal flexibility, the network rather deforms to prevent the fracture from occurring, as it observed in the snapshots inside Fig. \ref{fig:fracture}.  At large strain (from 20\%), only one molecular chain made of calcium and oxygen atoms still exists between the up and down phases. The latter eventually breaks as soon as the strain reaches 26\%.

The ductile behavior that is observed requires an extra care: indeed, as the crack propagates, irreversible processes, such as plasticity, occur inside a process zone around the crack tip. An estimated length of this plasticity zone $r_{{\rm pl}}$ can be evaluated using the Dugdale--Barenblatt formula \cite{dugdale_yielding_1960, barenblatt_mathematical_1962, lemm_advances_1962}:

\begin{equation}
 r_{{\rm pl}} = \frac{\pi}{8} \left( \frac{K_{Ic}}{\sigma_{{\rm pl}}} \right)^2
\end{equation} where $\sigma_{{\rm pl}}$ is the plastic yield stress of the material. In the present C--S--H sample, we find $r_{{\rm pl}}$ = 13.7 \AA, which is significantly larger than the value found in quartz (3.4 \AA\ \cite{brochard_fracture_2013}) but remains lower to what can be found in kerogen (19.9 \AA\ \cite{brochard_fracture_2013}). At the end of the fracture, the process zones located at both sides of the crack eventually overlap because of the periodic boundary conditions. As suggested in Ref. \cite{brochard_fracture_2013}, this feature can be taken into account by replacing in Eq. \ref{eq:integration} the real crack area $\Delta A _{\infty}$ by an effective area given by $\Delta A _{\infty , {\rm eff}} = \Delta A _{\infty} - L_x r_{{\rm pl}}/2$.

To get an estimation of the precision of the computed fracture toughness, five additional fracture simulations with different box and initial crack lengths were run. Even if each stress--strain curve is specific to each considered system, the resulting fracture toughness values remain consistent. This allows us to evaluate the standard deviations of the fracture properties as being $G_c = 1.72 \pm 0.29$ J/m$^2$ and $K_{Ic} = 0.369 \pm 0.030$ MPa.m$^{1/2}$.

Although, to the best of our knowledge, no experimental value of the fracture toughness of C--S--H at the scale of the grain is presently available, this value can be compared with available measurements in cement paste. Depending on the method used, the water-to-cement ratio and the age of the paste, the fracture toughness of cement is usually found between 0.29 and 0.40 MPa.m$^{1/2}$ \cite{dwivedi_strength_2013,nadeau_slow_1974,higgins_fracture_1976,hillemeier_fracture_1977,brown_fracture_1973}. Although our values, obtained at the nanoscale, cannot be directly compared with experimental results, obtained at much larger scales, the fact that the values are of the same order supports the ability of the present simulations to provide realistic results. We also note that we observed lower fracture toughness than in pure silica (around 0.8 MPa.m$^{1/2}$), which is consistent with the fact that water molecules depolymerize the silicate chains in C--S--H.

\begin{figure}
\includegraphics*[height=\linewidth, angle=-90, keepaspectratio=true, draft=\ddst]{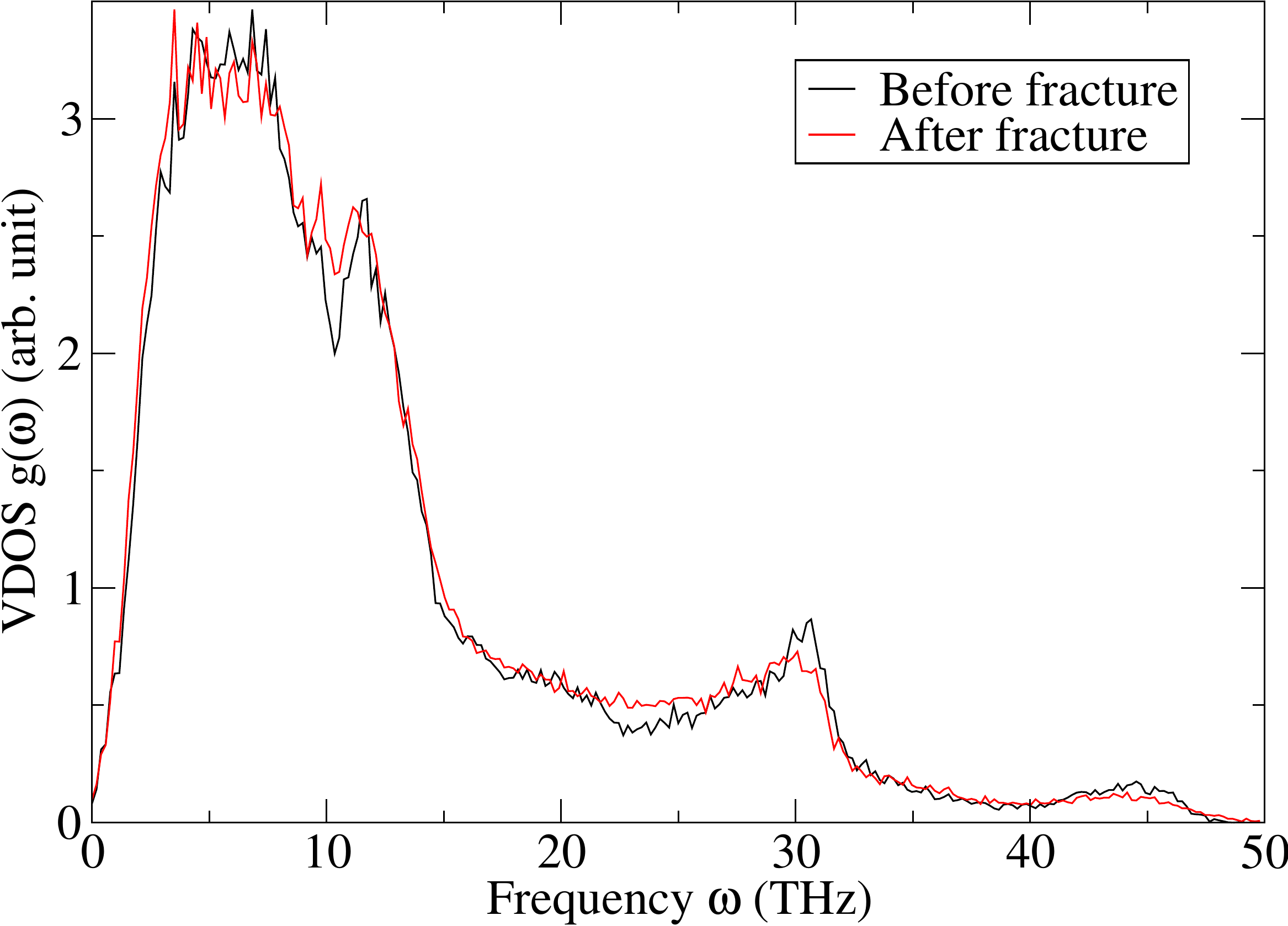}
\caption{\label{fig:vdos} (Color online) Vibrational density of state (VDOS) of C--S--H at 300K, both before (black curve) and after (red curve) the full propagation of the crack in the box.}
\end{figure}

\textit{A posteriori}, we check that there is no contribution to the fracture energy of the phonons in the equilibrated system after fracture. To this end, we compute the vibrational density of state (VDOS) $g( \omega )$ before and after the full propagation of the crack. This is achieved by calculating the Fourier-transform of the velocity autocorrelation function (VAF) \cite{bauchy_structural_2012}:

\begin{equation}
 g( \omega ) = \sum \limits_{j=1}^N m_j \int_{-\infty}^{\infty} {\rm VAF}_j(t) \exp ({\rm i}\omega t)\, \mathrm dt
\end{equation} where VAF$_j(t)$ is the VAF for atomic species $j$:

\begin{equation}
 {\rm VAF}_j(t) = \frac{1}{N k_B T} <\textbf{v}_j(t) \textbf{v}_j(0)>
\end{equation} and where $N$ is the number of atoms, $m_j$ is the mass of an atom $j$,  $\omega$ is the frequency, and $\textbf{v}_j(t)$ is the velocity of an atom $j$. Fig. \ref{fig:vdos} shows the VDOS of C--S--H, both before and after fracture. Unfortunately, no experimental VDOS is currently available for C--S--H. However, we note that it is very similar to that typically observed in silicate glasses \cite{bauchy_structural_2012}, but features additional peaks at high frequency due to the H--O bonds. As expected, the VDOS does not change significantly after the fracture has happened. This is not surprising since, as mentioned before, the system is always at equilibrium with a thermostat during the fracture process in the present methodology.

\section{Discussion}
\label{sec:dis}

\begin{figure}
\includegraphics*[width=\linewidth, keepaspectratio=true, draft=\ddst]{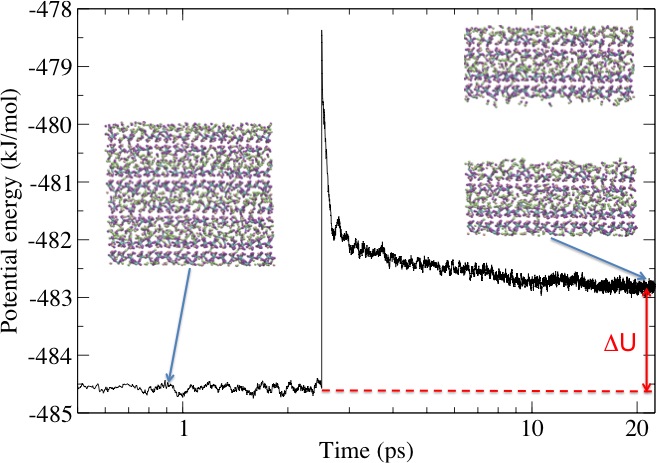}
\caption{\label{fig:surface} (Color online) Potential energy of C--S--H as a function of time. At $t=2.5$ ps, the system is cut into two parts.}
\end{figure}

We now aim to quantify the brittleness of C--S--H at the atomic scale. The critical energy release rate $G_c$ can be expressed from the surface energy $\gamma_s$:

\begin{equation}
 G_c = 2\gamma_s + G_{{\rm diss}}
\end{equation} where $G_{{\rm diss}}$ captures all forms of dissipated energy linked to irreversible processes and is equal to zero for a perfectly brittle material. The surface energy $\gamma_s$ was roughly estimated from molecular dynamics simulation by cutting  the system into two parts, letting it relax for 25 ps, and computing the change of the potential energy of the system. For the present C--S--H sample, we find $2\gamma_s$ = 1.06 J/m$^2$. This value is fairly comparable to the surface energy of tobermorite gel, which is $2\gamma_s$ = 0.8 J/m$^2$ \cite{brunauer_surfaces_1965}. Note that we observe the dissociation of some free water molecules during the fracture. As, e.g., Si--O--Ca bonds break, terminal 1-fold O and under-coordinated Ca atoms are energetically unfavorable: therefore, the following chemical reaction occurs:

\begin{equation}
 \begin{split}
  {\rm (-Si-O-Ca-) + H_2O} \rightarrow  \\
   {\rm(-Si-O-H) + (H-O-Ca-)}
 \end{split}
\end{equation} This reaction obviously stabilizes the newly created surface, thus decreasing the surface energy. This stabilization can be observed in Fig. \ref{fig:surface} as it manifests by a gradual decrease of the potential energy after the fracture. This highlights that, to study the fracture of such a hydrated material by molecular dynamics, it is critical to use a reactive potential that can handle such chemical reactions. The use of a classic non-reactive potential would result in an overestimation of the fracture energy.

The knowledge of the surface energy allows estimating $G_{{\rm diss}}$=0.66 J/m$^2$, as well as a lower bound of the fracture toughness $K_{Ic,{\rm min}}$, i.e., the value one would get if no energy were dissipated during the fracture process. We find $K_{Ic,{\rm min}}$ = 0.289 MPa.m$^{1/2}$. The analysis allows quantifying the ductility of the material by computing a brittleness parameter $B=2\gamma_s/G_c$, which is equal to 1 for a perfectly brittle material. Here, we find $B=0.62$, which highlights the fact that the fracture of C--S--H at the atomic scale involves ductility and, therefore, cannot be captured by methods that only rely on the theory of fracture in brittle materials. The energy dissipated during the fracture can arise from different phenomena, such as plasticity or crack blunting. This will be investigated in future works.

Through all these findings, the question remains if a ductile response is to be expected on larger scales. All the specimens simulated in this study have a plastic zone length extending approximately 14 \AA\ from the crack tip. The plastic zone, in the continuum sense, is a nonlinear zone where characterized by progressive softening \cite{bazant_fracture_1997}. If the plastic zone size is very small compared to all relative dimensions of the simulation, then it can be approximated as a point, in which the models developed in LEFM can apply. However, if the plastic zone is non-negligible compared to specimen sizes, the specimens undergo progressive softening damage (due to mechanisms such as collective rearrangement of atoms or bond breakages) and stable crack propagation, where, for every increment in crack length, a stable equilibrium is obtained \cite{bazant_fracture_1997}. The type of materials that have a fixed plastic zone length and exhibit a ductile behavior for small specimen sizes and a brittle behavior for large sizes are known as quasibrittle because their behavior is size dependent \cite{bazant_size_1984}. Methods exist which can extract fracture properties from such material behaviors \cite{hoover_comprehensive_2013-1, hoover_cohesive_2014, hoover_comprehensive_2013, bazant_scaling_2005}. Further investigation on bulk C--S--H is needed to determine if the brittleness of the specimen increases with system size.

\section{Conclusion}
\label{sec:ccl}

By using a realistic model of C--S--H as well as a molecular dynamics-based method allowing the capture of non-elastic effects, we computed the values of the surface energy, fracture toughness, and critical energy release rate of C--S--H grains, which are not directly accessible from experiments. At the atomic-scale, C--S--H appears to break in a ductile way, so that one cannot rely on LEFM-based methods. The intrinsic fracture toughness of C--S--H grains appears to be very close to that of the cement paste obtained by different experimental techniques. This suggests that the C--S--H grains play a major role in the fracture of cement.

It is now necessary to upscale this result to the meso- and, eventually, the macro-scales to enable direct comparisons with experiments as well as the design of a tougher material using a bottom-up approach. As a preliminary step, starting from the atomistic scale, one should study how the use of additives to C--S--H can result in tougher cement. This could be achieved by tunning the elastic modulus and/or by maximizing the ductility to dissipate more energy during the fracture.

\section*{Acknowledgments}

This work was supported by Schlumberger under an MIT-Schlumberger research collaboration and by the CSHub at MIT. This work has been partially carried out within the framework of the ICoME2 Labex (ANR-11-LABX-0053) and the A*MIDEX projects (ANR-11-IDEX-0001-02) cofunded by the French program "Investissements d'Avenir" which is managed by the ANR, the French National Research Agency.

\end{document}